\documentclass[aps,pre,twocolumn,groupedaddress,showpacs]{revtex4}
\usepackage{amssymb}
\usepackage{graphicx,color}
\definecolor{r}{rgb}{1,0,0}
\definecolor{g}{rgb}{0,1,0}
\definecolor{b}{rgb}{0,0,1}

\begin{document}


\title{Abrasion of flat rotating shapes}
\author{A. E. Roth$^1$, C. M. Marques$^2$, and D. J. Durian$^1$}
\affiliation{$^{1}$Department of Physics \& Astronomy, University of Pennsylvania, Philadelphia, PA 19104-6396, USA}
\affiliation{$^{2}$Institut Charles Sadron, 23 rue du Loess, BP 84047, F-67034 Strasbourg Cedex, France}

\date{\today}

\begin{abstract}
We report on the erosion of flat linoleum ``pebbles'' under steady rotation in a slurry of abrasive grit.  To quantify shape as a function of time, we develop a general method in which the pebble is photographed from multiple angles with respect to the grid of pixels in a digital camera.  This reduces digitization noise, and allows the local curvature of the contour to be computed with a controllable degree of uncertainty.  Several shape descriptors are then employed to follow the evolution of different initial shapes toward a circle, where abrasion halts.  The results are in good quantitative agreement with a simple model, where we propose that points along the contour move radially inward in proportion to the product of the radius and the derivative of radius with respect to angle.
\end{abstract}

\pacs{45.70.-n, 83.80.Nb, 91.60.-x, 02.60.Jh, 81.65.Ps}
%


\maketitle




Pebbles on a rocky beach or river bank are often flat, and exhibit a wide variety of smooth rounded forms.  This must arise from the combined effects of the initial pebble shapes, the material properties of the pebbles, and the entire history of erosion processes.  For Geology, an important issue would be to decipher this history from the observed collection of pebble shapes \cite{boggs}.  For Physics, an important issue would be to isolate and understand the physical action of different classes of erosion processes.  It is not known, for example, whether the variety of shapes in some actual set of pebbles reflects the initial conditions and the duration of an erosion process that would eventually produce perfectly circular pebbles.  Another possibility is that the responsible erosion process is stochastic, giving rise to a variety of shapes for any initial conditions.

Several models for the kinetics of two-dimensional pebble erosion have recently been proposed.  The simplest is a {\it ``polishing''} model, where the normal velocity of contour points is proportional to the local curvature and is zero where the curvature is negative \cite{PebblePRL}.  Under this action, any initial pebble shape approaches a circle in the limit of vanishing area \cite{grayson}.  This is similar in spirit to what might be called the {\it ``Aristotle''} model, where the velocity of contour points would be directed toward the center of mass and grow with radial distance \cite{aristotle}.   We are aware of no actual data that are explained by either of these models.  A stochastic {\it ``cutting''} model has also been proposed, where a straight cut is made from a random contour point with a length drawn from an exponential distribution \cite{PebblePRL}.  This model successfully captures some features of laboratory erosion of clay pebbles in a rotating tray.  However it is incapable of generating concave regions of negative curvature, which exist in the laboratory experiments and which may or not be important for natural erosion processes.  And more recently, an analytically tractable {\it ``chipping''} model has been proposed, where a randomly selected corner is broken off \cite{RednerPRE07}.  This model produces nontrivial anisotropic shapes.

Comparison of data to such models requires that shape be quantified.  In Geology, shapes are often described verbally (angular, rounded, elongated, platy) or by comparison to a standardized charts \cite{boggs}.  It is also common practice to construct dimensionless ratios from measured values of long vs intermediate vs short axes \cite{illenberger, benn, howard, hofmann, graham}.  To better connect with the microscopic action of erosion, other shape quantifiers have been constructed in terms of the curvature \cite{PebblePRL} or the turning angle \cite{RednerPRE07} at each point along the contour.  Intuitively, regions of high positive curvature are more exposed and hence subjected to faster erosion.  Unfortunately, as reviewed in the Appendix of Ref.~\cite{PebblePRE}, it remains difficult to reliably measure curvature from digital images because this involves numerical computation of a second derivative.

Thus it would be useful to explore a specific erosion process with reproducible deterministic action.  And it would be useful to establish reliable means for extracting curvature from digital images.  Towards these ends, we conduct experiments on the abrasion of soft flat shapes by rotation in a slurry of abrasive grit.  We show that the erosion is deterministic and reproducible, and gives rise to circular shapes of nonzero size.  This does not correspond to either the ``polishing'' or ``Aristotle" models, but can be described by another similarly simple evolution equation.  In addition we introduce a straightforward measurement procedure in which multiple digital photographs are taken at different orientations, in order to effectively average over pixelation noise.  We show that this permits the local curvature to be measured with an uncertainty that is purely statistical and of well-controlled magnitude.


\section{Materials and methods}

\subsection{General}

\begin{figure}
\includegraphics[width=3.0in]{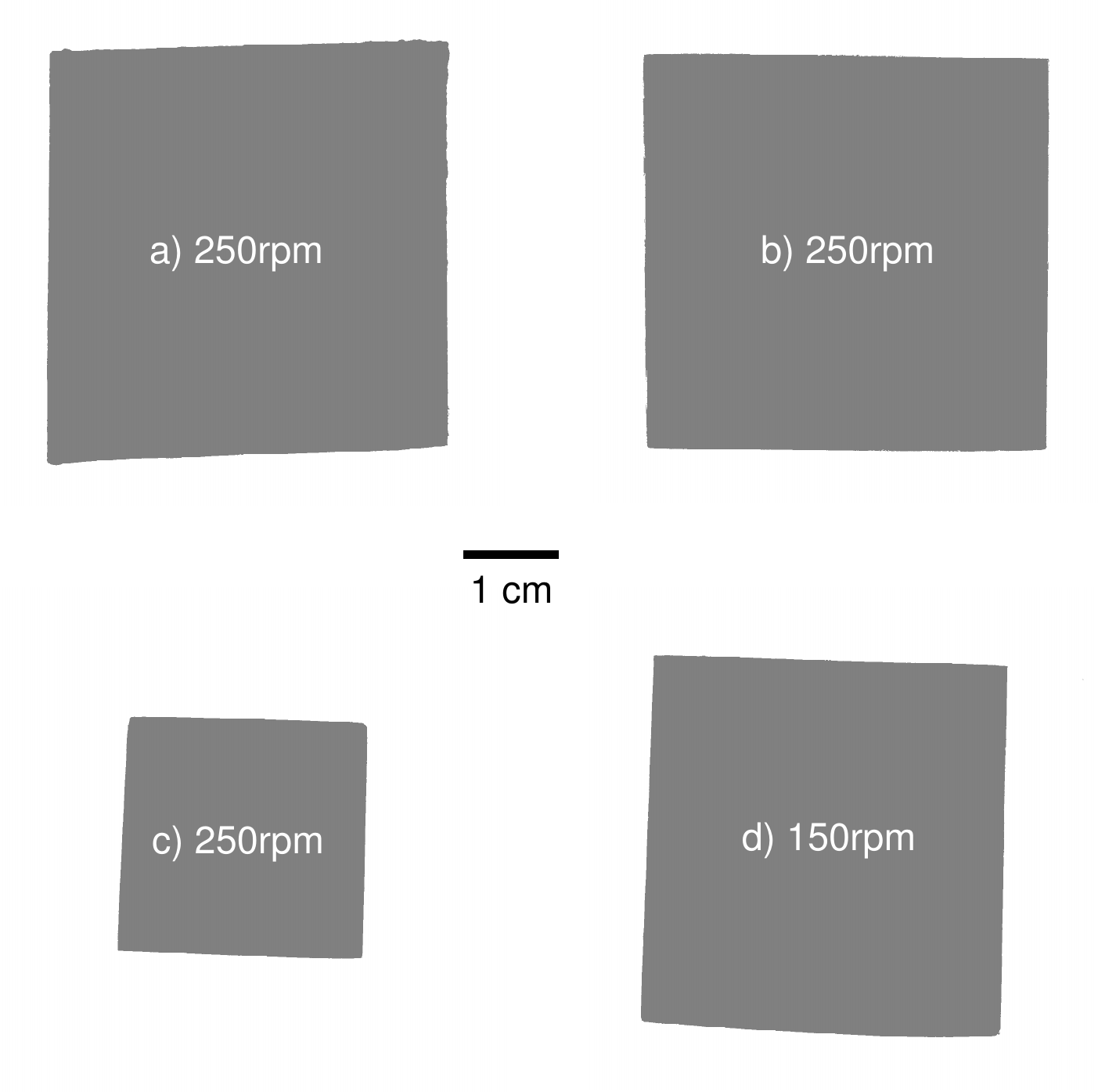}
\caption{Binarizied photographs of four flat linoleum squares, prior to erosion. Note that the construction process produces some level of variability and degree of negative curvature in the contours.  }
\label{negcurv}
\end{figure}

\begin{figure*}
\includegraphics[width=5.5in]{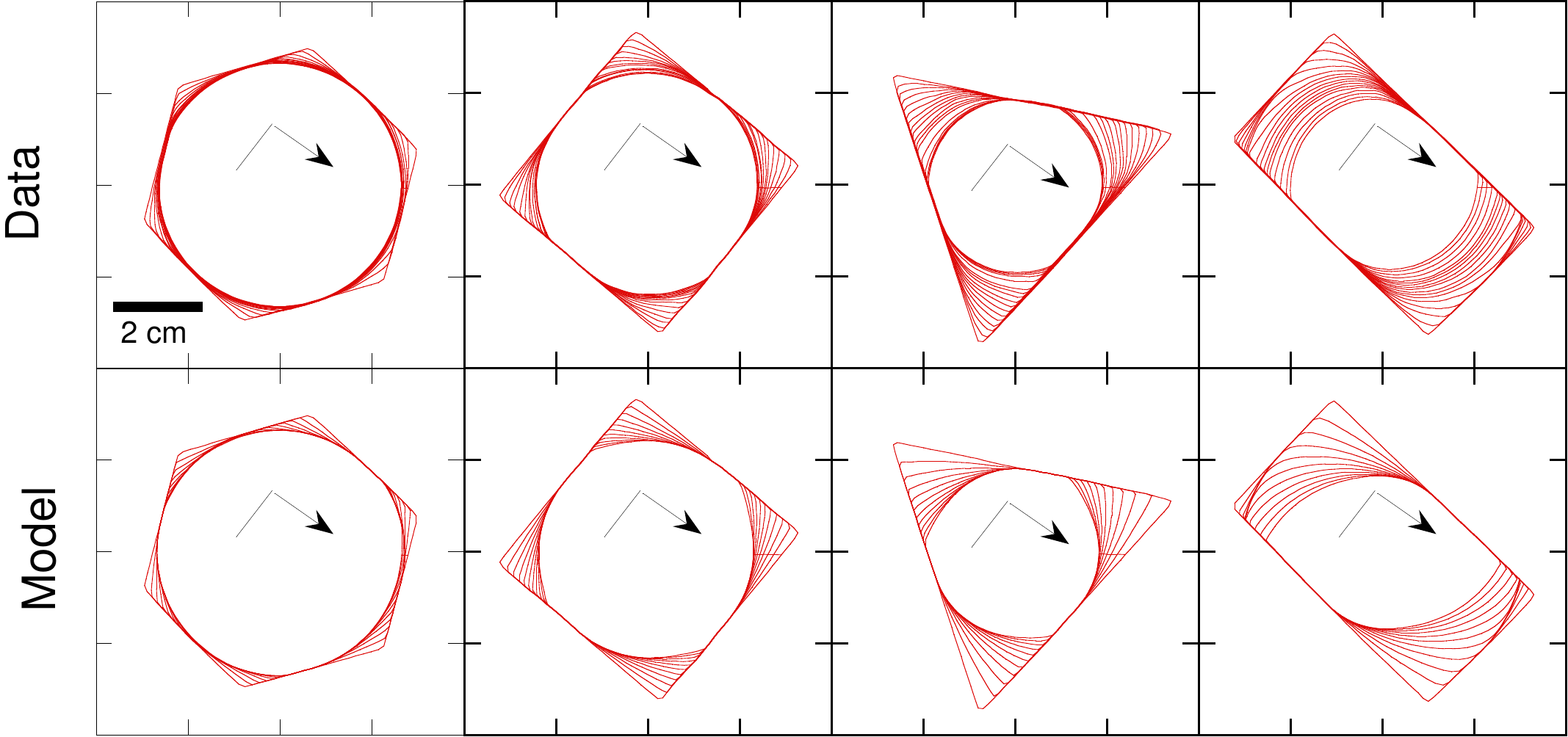}
\caption{(Color online) Contour sequences for linoleum pebbles abraded by clockwise rotation in a slurry of grit.  The top row shows experimental data at equal intervals of 7500 rotations; the bottom row shows the evolution of the initial contours under the action of Eq.~(\protect{\ref{model}}) with $\alpha=\beta=1$.  The square pebble is the one labeled (a) in Fig.~\protect{\ref{negcurv}}.  }
\label{contours}
\end{figure*}

In order to have a set of flat pebbles with uniform isotropic consistency, we choose linoleum tiles of thickness 1/8~inch (3.175~mm).  Linoleum is a  commercial floor covering made from renewable materials such as solidified linseed oil, pine rosin, ground cork dust, wood flour, and mineral fillers such as calcium carbonate.  The product we chose has no backing or fibrous content.  Initial shapes are formed with a standard tile cutter, and then filed down to remove surface texture.  These include four squares with approximate edge lengths of 2.5 and 5~cm; a hexagon with edge length 3~cm; a triangle with edge length 6~cm; and a $2.5\times 5~{\rm cm}^2$ rectangle.  Photographs of the squares are displayed in Fig.~\ref{negcurv}, converted from grayscale to binary.  Note that the tile cutter does not produce identical shapes, and that the edges all possess slight concave regions with small negative curvature.  A 6~mm mounting hole is drilled through the center of each shape.  An additional 3~mm fiducial marker hole is drilled 1~cm from the center, for determining the orientation of the shape in photographs.

Erosion is accomplished by rotation in a slurry of silicon carbide grit (16 mesh, McMaster-Carr product No. 4780A34) completely submerged in water.  The container holding the slurry has diameter of 12.5~cm, and is filled with grit to a depth of 10~cm; water covers the slurry by a few cm.  The grains have irregular shapes, an average size of $d=1.1$~mm, a polydispersity of about 50\%, a density of $\rho_g=3.21$~g/cc, and a packing fraction of about 60\%.  The pebbles are mounted by screw and lock-washer to a vertical steel rod attached to a Barnant series 20 mixer.  The pebbles are carefully lowered $H=5.5$~cm into the grit.  The rotation is exclusively clockwise at a rate of 250~rpm, except for one square where the rate is 150~rpm.  As erosion proceeds, linoleum debris floats to the surface of the water, where it is regularly skimmed off. 

At regular intervals the pebble is removed from the grit, laid on a lightbox, and photographed from directly above with a Nikon D70 six-megapixel digital camera equipped with a Nikon AF Micro Nikkor 60mm lens.  The magnification is such that pixels collect light from $0.04\times0.04~{\rm mm}^2$ regions on the pebble.  Images are converted to binary, and the skeletonized contour is identified, using built-in LabVIEW commands.  Example contours are shown for four shapes in Fig.~\ref{contours}.  Note that the erosion is chiral and is faster at leading edges, in accord with the clockwise sense of rotation.  Note also that the contour spacing decreases, showing that the abrasion slows down as the final circular shape is approached.  One convenient feature of our choice of system is that this process comes to completion within roughly one day.  Another convenient feature is that the grit is much harder than the pebbles, and does not change as the pebble erodes.

The flow of the slurry in response to pebble rotation, and its variation around the perimeter of the pebble, would be important for a first-principles model of the abrasion process.  Unfortunately, however, the grit is opaque so we cannot visualize the flow very well.   At the translucent wall of the container, some motion could be observed in the plane of the pebble with a height about 7 to 8~mm.  The rate of this flow decreases toward zero as the pebbles become circular.  The surface of the slurry always remains at rest.  Several dimensionless numbers help characterize the forces at play.  The first is the Reynolds number based on grain size $d$ and the speed $v$ at the perimeter, which also sets the scale for relative grain motion near the perimeter: Re=$ \rho_f v d/\eta \approx 10^3$, where $\rho_f$ and $\eta$ are the fluid density and viscosity respectively.   This means that the flow of the water at the edge of the pebble and also between the surrounding grains is mildly turbulent, such that the viscosity of the water plays no major role.  The corresponding Stokes number, for the ratio of grain inertial to fluid viscosity forces, is three times larger since the grains are three times denser than the water.  Another important number would be the ratio of grain inertial to friction forces, which can be estimated as $\rho_g v^2 / [ \mu (\rho_g-\rho_f)gH] \approx 2$, assuming a friction coefficient $\mu$ of order one.  In short, the fluid inertial, grain inertial, and friction forces are all comparable and much greater than viscous forces.

The remainder of this section concerns experimental details and is organized as follows.  The following two subsections describe our multiple photograph method for eliminating systematic pixelation errors in the contour location and for calculating statistical errors.  Then the final subsection reviews the shape descriptors to be employed for quantifying shape evolution by rotational abrasion and for comparing to a model in subsequent sections.


\subsection{Multiple photographs}

\begin{figure*}
\includegraphics[width=5.75in]{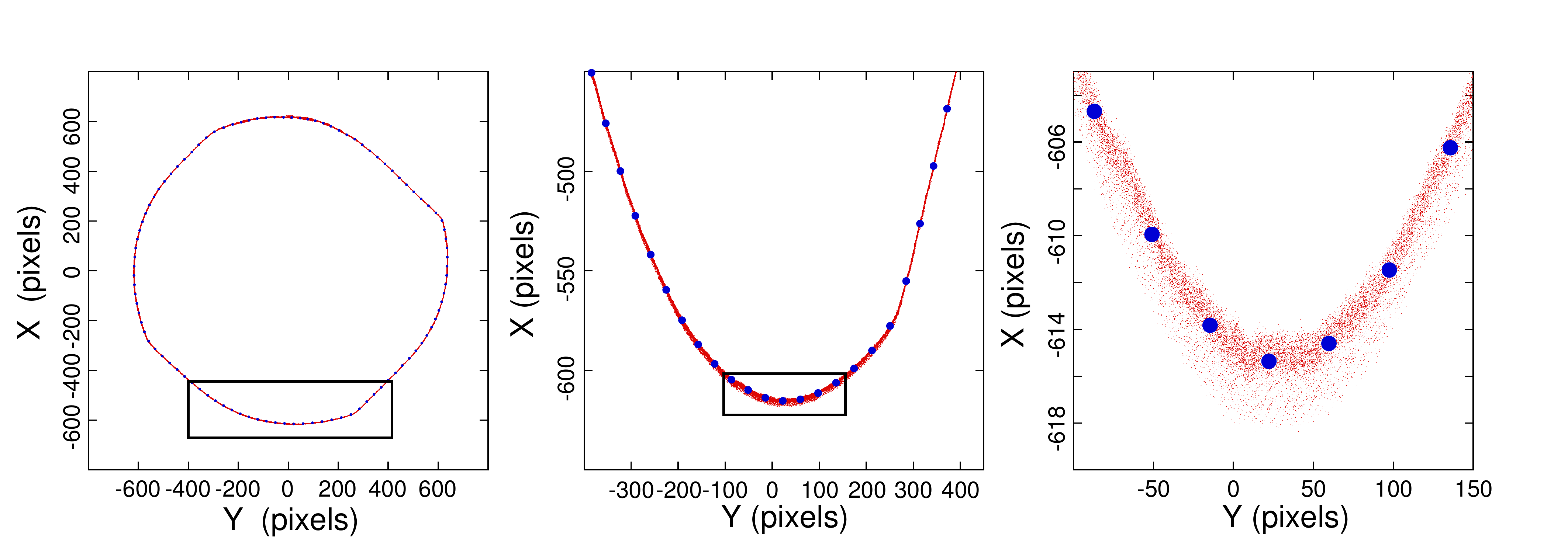}
\caption{(Color online) Cloud of pixel data (small red dots) and vertex points for the final polygonal contour (large solid blue circles), shown in pixel units (0.04~mm) at various levels of magnification.  The cloud consists of the skeletonized contours from 100 digitial photographs taken at different angles, with approximately 5000 pixel points per contour.}
\label{cloudpoints}
\end{figure*}

\begin{figure}
\includegraphics[width=3.00in]{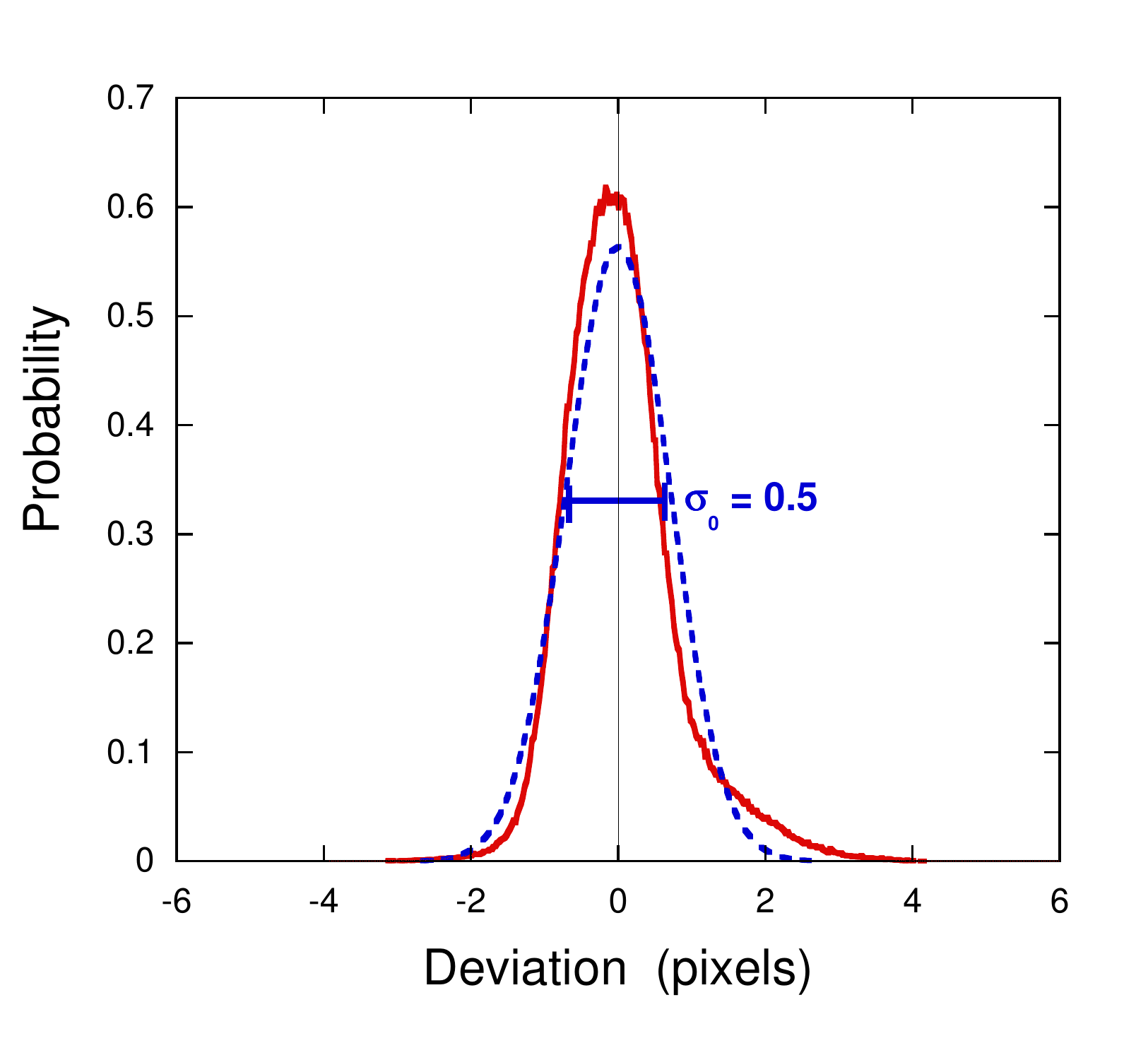}
\caption{(Color online) Normalized histogram of deviation of pixel points from contour.  Positive values are for points outside the contour and negative values are for points inside the contour.  The blue dashed curve depicts a Gaussian with standard deviation of 0.5 pixels.  }
\label{deviationvsangle}
\end{figure}

The skeletonized contour points given by analysis of a digital image is satisfactory only for computing the area and linear dimensions of the pebble.  Since the points are all on a square grid, neither the number of points nor the sum of distances between adjacent points give an accurate measure of the perimeter.   The difficulty is compounded for computing the unit tangent vector ${\bf T}$, and even more so for the curvature vector ${\bf K}={\rm d}{\bf T}/{\rm d}s$, where $s$ is arclength.  The approach taken in Refs.~\cite{PebblePRL, PebblePRE} was to fit radius vs angle to a cubic polynomial, averaging over a range of acceptable fitting windows.  Here we develop an alternative approach in which the pebble is photographed at multiple orientations with respect to the grid of pixels in the digital camera.  For this, the lightbox on which the pebble rests is placed on a rotation stage directly under the camera.  Both the camera and the stepper-motor for the stage are automated by LabVIEW to take 100 photographs at equal angle intervals over a range $0-\pi/4$.  Pixelized contour points are then aligned to a common coordinate system according to the location of the mounting and fiducial holes.  An example of the final cloud of raw pixel points is shown in Fig.~\ref{cloudpoints} at three levels of magnification, zooming from the entire contour down to the pixel scale.   Note that this pebble is about 1200 pixels across, and that the alignment of the multiple images is good to the pixel scale.  Also note that the pixel points cluster densely with only a little systematic structure.

The nature of the noise in the pixel points is investigated in Fig.~\ref{deviationvsangle} by a normalized histogram for the distance between pixel points and the estimated location of the actual contour.   Note that this distribution is approximately Gaussian, and has a standard deviation close to 1/2 pixel, $\sigma_o=\ell/2$.  Thus we may safely treat the pixel points as having random uncorrelated Gaussian noise.  By contrast, the uncertainties in adjacent points on a single skeletonized image are highly correlated, and can lead to unknown systematic errors in the computation of the local tangent.


\subsection{Polygonal contour and uncertainty}

\begin{figure}
\includegraphics[width=3.00in]{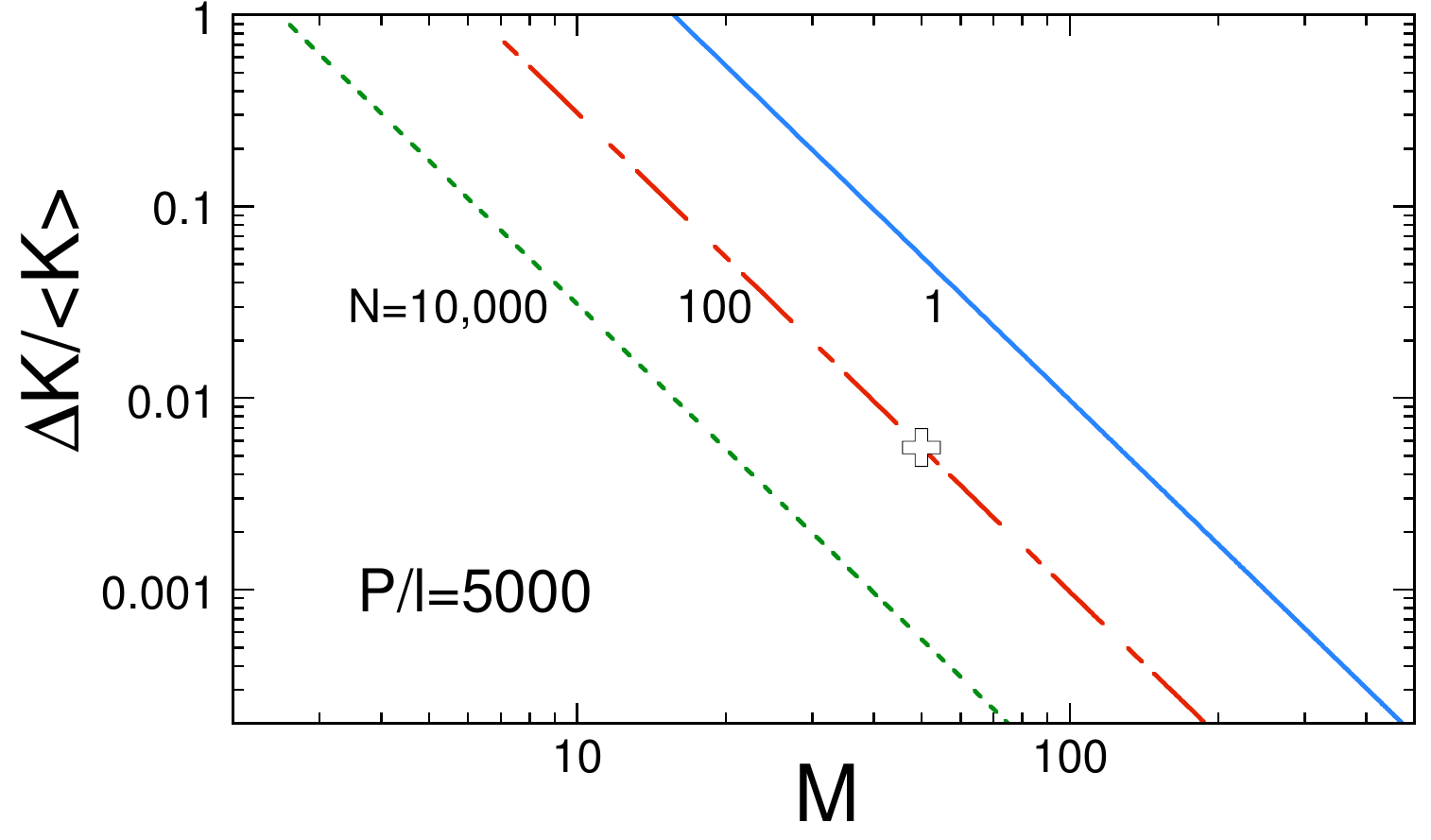}
\caption{(Color online)  Fractional uncertainty in curvature, Eq.~(\ref{DK2}), plotted vs the number $M$ of pixel points per photograph that are averaged together into a vertex point.  Examples are shown for a contour of perimeter $P/\ell=5000$ pixel units photographed at $N$ different angles, as labeled.  The open plus sign marks the combination of parameters used in our experiments.}
\label{curvuncertainty}
\end{figure}

The final step in treating the data is to construct a polygonal contour with roughly equal segment lengths based on the cloud of pixel points obtained from $N$ multiple images.  For this we begin by sorting all contours by angle.  Then we divide one of the contours into intervals with $M$ pixel points, and average all the points in each interval to create `seed' points for the vertices of the final polygon.  The position of each seed point is then refined by averaging together the closest $M$ pixel points from each of the $N$ images.  This process is repeated three times, which is sufficient for convergence.  The final vertex points for a polygonal contour are shown in Fig.~\ref{cloudpoints}, as solid blue circles, for the choice $M=50$.  In this example, there are $N=100$ images consisting of roughly 5000 pixels; therefore, there are roughly 100 evenly-spaced vertices in the final polygonal contour, each formed by the average of $NM=5000$ pixel points.  The choice $M=50$ is made so that the statistical uncertainty in the local curvature falls below the one percent level, as demonstrated next.

The statistical uncertainty of the local curvature may be estimated as follows, using the Fig.~\ref{deviationvsangle} result that each pixel point has a random Gaussian uncertainty of one half pixel size, $\sigma_o=\ell/2$.   First note that the curvature at a vertex is $K=\theta/S$ where $\theta$ is the turning angle between adjoining straight line segments of approximate length $S=M\ell$.  The uncertainty in $K$ is due entirely to turning angle uncertainty, which equals the uncertainty $\sigma_o/\sqrt{NM}$ in the vertex positions perpendicular to the contour divided by $S$.  The $\sqrt{NM}$ reduction assumes that the $NM$ pixel points per vertex are all uncorrelated.  Three vertices are involved in defining the bending angle, and this gives an additional factor of $\sqrt{6}$.  Normalizing by the average curvature $\langle K\rangle = 2\pi/P$, where $P$ is the perimeter, gives the estimated percent uncertainty in curvature as
\begin{eqnarray}
	{\Delta K \over \langle K\rangle} &=& {\left(\sqrt{6}  {\sigma_o\over S\sqrt{NM}} \right) \over S} {P\over 2\pi}, \label{DK1}\\
						&=& {\sqrt{6}\over 4\pi} {P/\ell \over \sqrt{NM^5}}.\label{DK2}
\end{eqnarray}
The term in round brackets in Eq.~(\ref{DK1}) is the uncertainty in turning angle, and the simplification to Eq.~(\ref{DK2}) was made using $\sigma_o=\ell/2$ and $S=M\ell$.  Thus the curvature uncertainty scales as the number $P/\ell$ of pixel points in the skeletonized image contours divided by $\sqrt{NM^5}$ where $N$ is the number of images and $M$ is the number of pixel points per image that contribute to each vertex of the final polygonal contour.

The Eq.~(\ref{DK2}) result for the fractional uncertainty in curvature, $\Delta K / \langle K\rangle$, is plotted in Fig.~\ref{curvuncertainty} vs the number $M$ of pixel points per image that are averaged together into vertex points for the final polygonal contour.  Here the value of $P/\ell$ was taken as 5000, which is the typical number of pixel points in skeletonized contours for a compact pebble that fills the field of view of a six megapixel digital camera such as ours.  For only one image, $N=1$, a window size of $M\approx20$ is needed for $\Delta K$ to be smaller than $\langle K\rangle$; this explains the difficulties and pains taken to deduce the curvature from polynomial fits in Refs.~\cite{PebblePRL, PebblePRE}.  For $N=100$ images, the curvature uncertainty falls below the one percent level for $M=50$, as denoted in Fig.~\ref{curvuncertainty} by a large open plus sign.  This corresponds to the experiments reported here, as illustrated by the solid blue vertex points in the example of Fig.~\ref{cloudpoints}.  While a larger choice for $M$ would reduce the uncertainty further, it would give fewer than 100 points in the final polygonal contour and the resulting straight-line segments would eventually begin to deviate from the cloud of pixel points.  Further reduction in curvature uncertainty could also be obtained by increasing the number $N$ of images taken.  To obtain an independent sampling of the contour against the grid of pixels, the minimum rotation increment between successive images should cause each contour point to move by at least one pixel; furthermore the maximum total rotation should be $\pi/4$.  Therefore, a hard upper limit on $N$ would be one-eight  $P/\ell$.  Our choice of $N=100$ is large enough for good statistics, but safely below this limit.  The statistical uncertainty in other quantities, such as perimeter and area, could also be estimated; these will be significantly less than $\Delta K/ \langle K\rangle$ since curvature computation involves differentiation.


\subsection{Shape descriptors}

\begin{figure}
\includegraphics[width=3.00in]{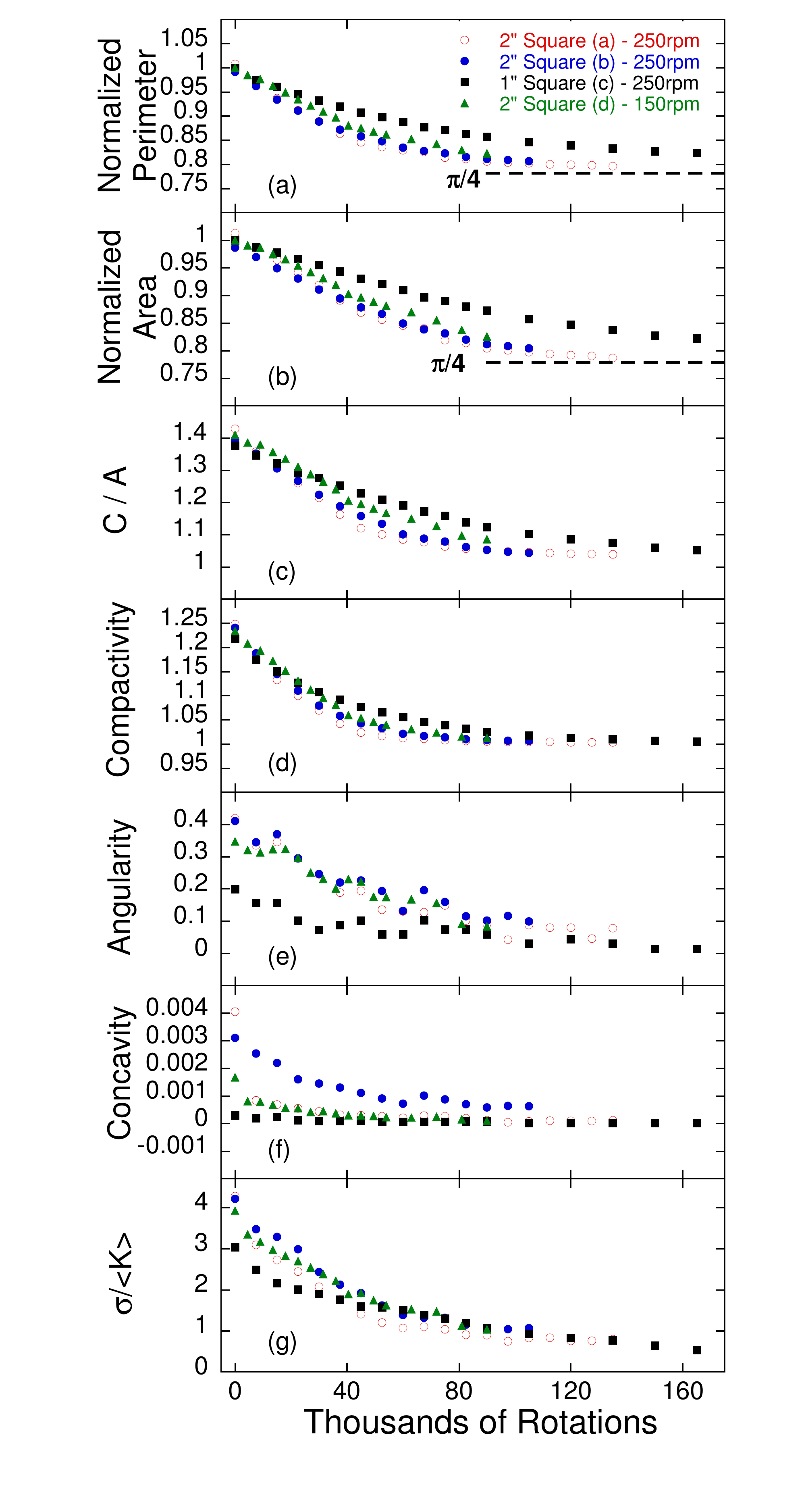}
\caption{(Color online) Dimensionless shape descriptors vs number of rotations, for the four different square pebbles pictured in Fig.~\protect{\ref{negcurv}}.  Consistent with evolution toward the largest inscribed circle, the normalized perimeter and area both approach $\pi/4$, the caliper aspect ratio $C/A$ and compactivity both approach 1, and the angularity, concavity, and width $\sigma / \langle K \rangle$ of the curvature distribution all decrease toward zero.  }
\label{squares}
\end{figure}

The concept of ``shape'' is somewhat nebulous and subjective.  We choose to quantify it using several different descriptors, all of which are demonstrated in Fig.~\ref{squares} showing evolution vs number of rotations in the abrasive slurry of grit for the four linoleum squares pictured in Fig.~\ref{negcurv}.  The first two shape descriptors are simply the perimeter $P$ and the area $A$, both normalized by their initial values.  The second two are the caliper aspect ratio $C/A$ and the compactivity.  The caliper aspect ratio is the ratio of the largest to smallest values measured by a caliper as the pebble is rotated.  The compactivity is a standard measure of circularity, equal to $P^{2} / (4\pi A)$.  Note in Fig.~\ref{squares} that all four of these measures are consistent with the pebble evolving from a square to the largest inscribed circle, for which the normalized perimeter and area both decay from 1 to $\pi/4$, the caliper aspect ratio decays from $\sqrt{2}$ to 1, and the compactivity decays from  $4/\pi$ to 1.

The remaining three shape descriptors shown in Fig.~\ref{squares} are all based on the curvature, measured at each vertex of the polygon as the turning angle per segment length.  The simplest is the ``angularity'', which we define as the fraction of the perimeter with negative curvature.  This quantifies a similar notion found in textbooks \cite{boggs}.  Next is the ``compactivity'', which is a standard quantity defined as the difference in area between the convex hull \cite{ConvexHull} and the actual shape, divided by the area of the actual shape.  Both the angularity and the concavity are zero for a shape that is purely convex.  And last is the width $\sigma/\langle K\rangle$ of the curvature distribution around the perimeter, divided by the average curvature.   This quantity, along with the cumulative distribution function of the curvature, were used in Refs.~\cite{PebblePRL,PebblePRE}.  For the four squares, the initial angularity is large but the compactivity is small, consistent with the small wavy imperfections and regions of slight negative curvature seen in the images of Fig.~\ref{negcurv}.  Note that the angularity, the concavity, and the width of the curvature distribution all decrease toward zero as the the squares abrade into circles.   Note also that of all the shape descriptors, $\sigma/\langle K\rangle$ is farthest from its asymptotic value when the experiment was stopped; the curvature distribution is thus most sensitive in detecting the unabraded flat regions seen by eye in the image sequences of Fig.~\ref{contours}.


\section{Shape Evolution}

\begin{figure*}
\includegraphics[width=6.0in]{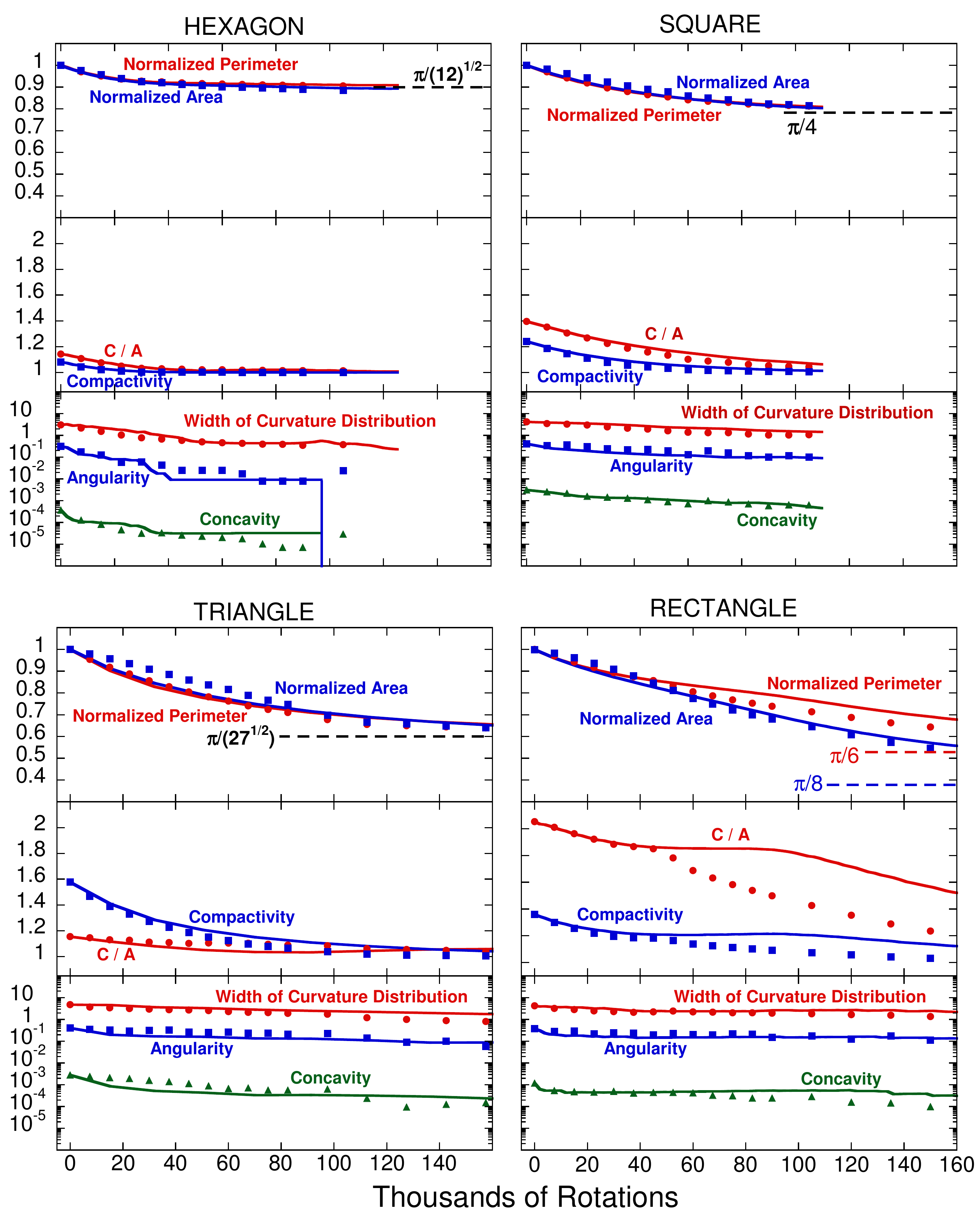}
\caption{(Color online) Dimensionless shape descriptors vs number of rotations, for four different initial shapes.  The top plots show size reduction; the middle plots show difference from a circle; the bottom plots show curvature-based descriptors.  Data are displayed as discrete points; the model Eq.~(\protect{\ref{model}}) is displayed as solid curves.  If the final shape is the largest inscribed circle, then data in the top plots should approach the dashed lines with indicated values, data in the middle plots should approach 1, and data in the bottom plots should approach 0.  Note that the range of each plot type is kept constant to better contrast the behavior of the different initial shapes. The square pebble is the one labeled (a) in Fig.~\protect{\ref{negcurv}}. }
\label{stats}
\end{figure*}

Using the above procedures, we now analyze image data in terms of shape descriptors in order to quantify the evolution of the initial shapes toward final circular shapes.  The qualitative evolution was already seen in Fig.~\ref{contours} for different initial shapes, and the quantitative evolution was already seen in Fig.~\ref{squares} for four squares.  The latter includes two nominally two-inch squares rotated at 250 rpm.  Comparison of their respective shape descriptors shows a fair degree of reproducibility, both in initial shape details and also in evolution.  Fig.~\ref{squares} also includes a nominal two-inch square rotated more slowly, at 150 rpm.  All its shape descriptors agree reasonably well with those for the faster abrasion, when plotted vs number of rotations rather than vs time.  The same holds for a nominal one-inch square rotated at 250 rpm, though in this case the initial shape is closer to a perfect square and the perimeter, area, and caliper aspect ratio all approach their asymptotic values a bit more slowly than the other shape descriptors.  Altogether, these observations show that abrasion by rotation in a slurry of grit is essentially independent of size and rate, and hence is controlled by geometry and materials properties alone.

In Fig.~\ref{stats} we display the evolution of all the shape descriptors for the square, hexagon, rectangle, and triangle, whose contour sequences are depicted Fig~\ref{contours}.  For all four shapes, the top plots in Fig.~\ref{stats} show the normalized area and perimeter, plus the asymptotic values for the largest inscribed circle; the middle plots show the caliper aspect ratio and compactivity, which both asymptote to one; the bottom plots show the width of the curvature distribution, the angularity, and the concavity, which all asymptote to zero for the largest inscribed circle.  Note that initial shapes that are closer to a circle decay more rapidly toward the final shape.  In particular, the $1/e$ decay constants for the normalized areas and perimeters are approximately 30K revolutions for the hexagon, 60K revolutions for the square, 90K revolutions for the triangle, and 160K revolutions for the rectangle.


\section{Rotational Abrasion Model}

\begin{figure}
\includegraphics[width=2.00in]{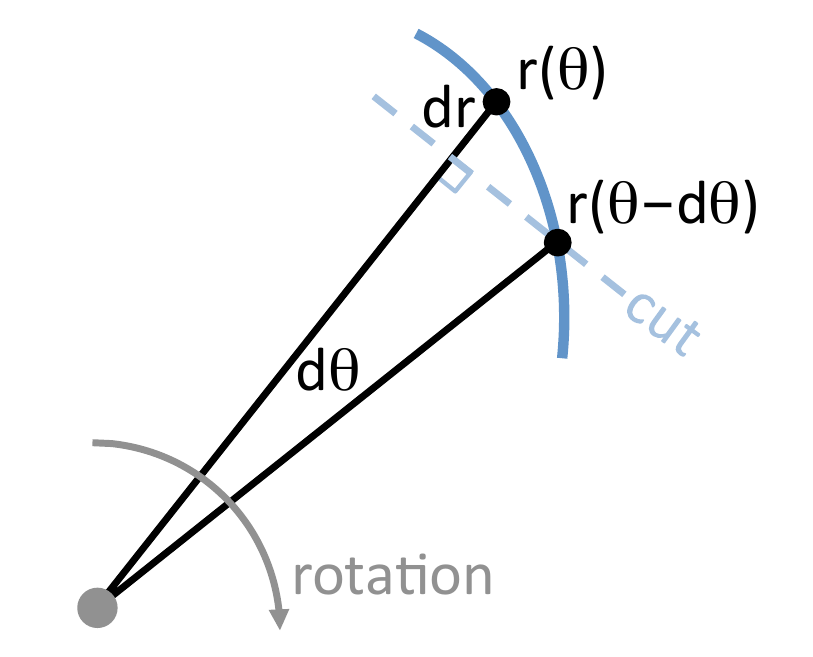}
\caption{(Color online) Schematic illustration of a geometric ``cutting'' model.  Due to rotational motion, material is removed by cuts normal to the radial direction: $|{\rm d}r|  = r(\theta)-r(\theta-{\rm d}\theta)\cos({\rm d}\theta)$. }
\label{cutting}
\end{figure}

In this final section we attempt to model the abrasion processes and compare with quantitative shape data.  Since the abrasion is due to rotation, we seek the rate of change of the radial coordinates $r$ of the vertex points.  For the first ingredient, in accord with Aristotle \cite{aristotle}, we suppose that the erosion is faster for points farther from the rotation axis, in proportional to a power of the tangential speed $(\omega r)^\alpha$.  For the second ingredient, we consider the extent to which a segment moves into the slurry.  This is determined by the magnitude and sign of the derivative of radius vs angle, ${\rm d}r/{\rm d}\theta$.  If zero or negative, there is no abrasion since the segment moves parallel to or away from the slurry.  The greater the positive magnitude, the more the segment penetrates into the slurry during rotation.  Altogether, we thus propose the rate of change of vertex radii to be
\begin{equation}
\frac{{\rm d}r}{{\rm d}t} \propto \left\{ \begin{array}{ll}  -r^{\alpha} (\frac{{\rm d}r}{{\rm d}\theta})^{\beta} &\quad {\rm d}r/{\rm d}\theta > 0 \\ 0 &\quad \textrm{otherwise} \end{array} \right.
\label{model}
\end{equation}
For any positive values of the powers $\alpha$ and $\beta$ this model gives abrasion that halts as the shape approaches a circle, where the radius is constant independent of $\theta$ around the entire contour.  Note, however, that this model becomes unphysical for pebbles where $r(\theta)$ is not 
single-valued.  For shapes far from a circle, where $r$ varies greatly with $\theta$, higher-order derivatives as well as non-local effects could become important.

The evolution of a given set $\{ r_i \}$ of vertex radii under Eq.~(\ref{model}) may be found by finite differencing as follows.  At each time step, points with $r_i>r_{i-1}$ are incremented by
\begin{equation}
{\rm d}r_i = -\Delta r_{max}  \left( { r_i \over \Delta r_{max} } \right)^\alpha
					\left( {r_i - r_{i-1} \over \Delta r_{max}}  {\Delta\theta_{min}\over \theta_i-\theta_{i-1}}  \right)^\beta,
\label{dr}
\end{equation}
where $\Delta r_{max}$ is the largest difference $r_i-r_{i-1}$, and $\Delta\theta_{min}$ is the smallest difference $\theta_i-\theta_{i-1}$.  This
corresponds to a variable time step of
\begin{equation}
{\rm d}t \propto \Delta r_{max}  \left( { 1 \over \Delta r_{max} } \right)^\alpha
					\left(  {\Delta\theta_{min}\over \Delta r_{max} }  \right)^\beta,
\label{dt}
\end{equation}
so that the ratio of Eq.~(\ref{dr}) to (\ref{dt}) gives Eq.~(\ref{model}).  This time step is sufficiently small by construction, as  confirmed by repeating with even smaller time steps.  While the model is not linear, analytic solution has been achieved; see companion paper~\cite{Bryan}.

The pebble evolution given by Eq.~(\ref{model}) for the simplest choice of $\alpha=1$ and $\beta=1$ is depicted qualitatively by contours in Figs.~\ref{contours} and quantitatively by the shape descriptors in Fig.~\ref{stats}.  In these figures the agreement with actual data for the hexagon, square, and triangle is very good.  For the rectangle, the agreement is satisfactory at early stages but becomes less so at later times.  On the other hand, good agreement is found if the model is initiated with a later-stage contour that is more compact.  For all four shapes the {\it same} proportionality constant was used in Eq.~(\ref{model}), as determined by matching the $1/e$ decay constant for the hexagon.  We note that similar degrees of agreement are found by fixing $\beta=1$ and taking $\alpha$ as 1/2, 1, 2, or 3; thus the model is relatively insensitive to the value of $\alpha$, which we thus take as 1 for simplicity.  By contrast, poor agreement is found by fixing $\alpha=1$ and taking $\beta$ as 1/2 or 2.
 
The observations $\alpha=\beta=1$ can be understood as follows in terms of a geometric cutting model. Given two consecutive angles $\theta-d\theta$, $\theta$  and the associated radii values $r(\theta-d\theta,t)$,  $r(\theta,t)$ at time $t$, one can compute the time evolution of $r(\theta,t)$ under a microscopic cut of the profile. In our case of rotating pebbles, the cutting forces act normally to the radial direction, as displayed in  Fig.~\ref{cutting}. One thus has ${\rm d}r = r(\theta,t+{\rm d}t) - r(\theta,t) = - [r(\theta,t) - r(\theta-{\rm d}\theta,t) \cos({\rm d}\theta)]$, and taking the limit of continuous variables gives ${\rm d}r/{\rm d}t= - w\ {\rm d}r/{\rm d}\theta$, where $w$ is the fraction of angle removed by unit time. Note that the dependence of ${\rm d}r/{\rm d}t$ on ${\rm d}r/{\rm d}\theta$,  is a direct consequence of the assumed tangential orientation of the cuts: a value of  $\beta=1$ is imposed in our experiments by the rotation geometry. The parameter $w$ carries information on the length and frequency of each successive microscopic cut, which is a function, for a given material and abrasion agent, of the tangential velocity only. One would thus expect $w$ to be proportional to the radius $r$, compatible with a value $\alpha=1$ in the model presented above.

\section{Conclusion}

In summary we have developed a method for measuring the contour of flat pebbles using multiple photographs from different angles.  This method produces accurate contours, and allows curvature to be deduced with a known degree of statistical uncertainty without systematic error.  It is our hope that this general procedure will be broadly applicable to research involving shape quantification, in the field and in the lab.  Using this advance, we have explored an erosion process where abrasion is caused by steady rotation in a slurry of grit.  By comparing different size squares and different rotation speeds, we found that the sequence of shapes evolves deterministically toward the largest inscribed circle.  By comparing different initial shapes, we have found that those closest to a circle approach the limiting shape more rapidly.  We have successfully modeled this behavior quantitatively with a simple differential equation, where contour points move radially inward in proportion to radius and the derivative of radius with respect to angle.  This model is different from both the deterministic ``polishing'' and ``Aristotle'' models, and is the only deterministic model of which we are aware that accounts for actual data.  It is our hope that these models may serve as a starting point for future theories of stochastic erosion, perhaps by the addition of a noise term, in order to compare with natural erosion processes.

\vfil

\begin{acknowledgments}
We thank B.G. Chen for helpful conversations.  A.E.R. thanks Roy and Diana Vagelos for their Science Challenge Award undergraduate scholarship.  This work was supported by the National Science Foundation through grant DMR-0704147.
\end{acknowledgments}

\bibliography{PebbleRefs}

\begin{thebibliography}{13}
\bibitem[{Con()}]{ConvexHull}
\bibinfo{note}{The convex hull is the boundary of the smallest convex region
  that contains the entire pebble, and can be visualized by imagining a rubber
  band stretched tight around the pebble.}
\end{thebibliography}

\end{document}